\documentclass{article}            
\usepackage{latexsym,amsfonts,amssymb,proof,epsf}
\usepackage[dvips]{graphics}

\newcommand{\ket}[1]{|{#1}\rangle} 
\begin{document}

\title{{
\bf Logical Interpretation of a Reversible Measurement\\
in Quantum Computing
}}

\author
{
Giulia Battilotti and Paola Zizzi
\\
\\
Dipartimento di Matematica Pura ed Applicata
\\
Universit\`a di Padova
\\
via G. Belzoni n.7, I--35131 Padova, Italy
\\
\\
giulia@math.unipd.it
\\
zizzi@math.unipd.it
}
\date{}
\maketitle

\begin{abstract}
We give the logical description of a new kind of quantum measurement that is a reversible 
operation performed by a hypothetical insider observer, or, which is the same, a quantum 
measurement made in a quantum space background, like the fuzzy sphere.
The result is that the non-contradiction and the excluded middle principles are both 
invalidated, leading to a paraconsistent, symmetric logic. Our conjecture is that, 
in this setting, one can develop the adequate logic of quantum computing. The role of 
standard quantum logic is then confined to describe the projective measurement scheme. 
 \end{abstract}

\section{Introduction}
Since a quite long time, it was assumed that standard quantum logic 
\cite{BvN} was the right logic for the quantum world. 
However, standard quantum logic fails when trying to describe a closed quantum system, 
like a quantum computer during the computational process. In the context of quantum computing 
\cite{}, standard quantum logic is only able to describe the standard 
(projective) quantum measurement, not the whole computational process (which in fact looks like 
a ``black box" to an external observer).
This ``fallacy" of standard quantum logic has been already recognized in the literature, as in 
\cite{DC}, where paraconsistency \cite{P} and linearity \cite{G} play relevant 
roles.
Also Basic Logic \cite{SBF} can be considered a promising alternative to standard 
quantum logic, in the context of quantum computing.
The main aim of our work is to look for a logic which describes the whole quantum computational 
process, comprising the measurement, from ``inside". 
The first step we do in this direction is to illustrate, in logical terms, a reversible 
quantum measurement, with no hidden quantum information, performed by a hypothetical ``insider 
observer" \cite{Z}. In the literature, the problem of a reversible quantum measurement has 
been considered, up to now, from a true physical point of view, see for example: 
\cite{MZ}; \cite{NCC}; \cite{NCCSB}; \cite{Mas}. Instead, our 
kind of reversible quantum measurement is a purely theoretical tool to investigate the internal 
computational state. The ``insider observer" is a fictitious being, who is used to describe the 
quantum measurement scheme in a quantum space-background. Actually, we discuss three different 
kinds of reversible measurements applied to one qubit: the mirror measurement, which does not 
change the probabilities, the fuzzy measurement, which mixes up the probabilities, and the Liar 
measurement, which interchanges the probabilities. The first one is conceived in a discrete, 
but still classical space, a ordered 2-points lattice, embedded in the fuzzy sphere \cite{Mad}
 with two cells, and corresponds to an anti-clockwise rotation about the z-axis of the Bloch 
sphere. The second one is conceived in the fuzzy sphere with two cells, and corresponds to a 
generic rotation of the Bloch sphere. The third one, is conceived in the same 2-points lattice 
as the first one, but with the two points interchanged. 
Then, our reversible quantum measurement is a kind of  ``thought experiment", which is 
useful to tune our reasoning with the internal logic of quantum computation.
Moreover, this new kind of quantum measurement might give some fresh insights in the 
foundations of quantum mechanics. In fact, the reversible measurement performed in this way, 
offers an interpretation of quantum mechanics which is very much on line with that of Mermin 
\cite{Mer}, as it attributes objectivity to the probabilities of the superposed state of one 
qubit, and separates this objective reality from the external observer and his knowledge.
This theoretical construction has been used to build up a computational model for Loop 
Quantum Gravity and quantum black holes \cite{Z2}. Moreover, it lead to a new 
approach to quantum computability and to a quantum interpretation of Godel's first incompleteness theorem
\cite{Z3}.
A preliminary treatment of the logical aspects arising from the reversible quantum measurement 
scheme, can be found in \cite{B}. 
In this paper, we develop the arguments introduced in \cite{Z} and \cite{B} and 
we illustrate the geometrical origin of logical symmetry which is due to the symmetry between 
the mirror measurement and the liar measurement. Also, we show how the logic of the insider 
observer is related to the standard quantum logic of the external observer, and to classical 
logic.
Here, as in \cite{Z} and \cite{B}, we consider the toy-model of one qubit. 
Due to the reversible measurement scheme, we get two new axioms (symmetric to each other) 
which are the opposite of the excluded middle and the non-contradiction principles. Thus, 
the logic of the insider observer is paraconsistent and symmetric.
However, if an external observer performs a standard quantum measurement, the excluded middle 
and the non-contradiction principles can be recovered. 
This paper is organized as follows: In Sect.2, we illustrate the reversible quantum measurement, 
more specifically, the mirror measurement, the fuzzy measurement, and the Liar measurement; in Sect. 3 
we describe the logic of the mirror and of the Liar measurements and we discuss some philosophical implications 
of the mirror measurement; in Sect. 4, we illustrate, in logical terms, the border of the black box, which is 
equivalent to the standard (projective) quantum measurement. Sect. 5 is devoted to the conclusions.  

\section{The Reversible Measurement Scheme}\label{misure}
As it is well known, the interpretational problem of quantum measurement, when performed by an 
external observer, is one of the hardest in the foundations of Quantum Mechanics. However, we 
believe that it is not a problem due to our lack of understanding (or knowledge), but it is just a consequence of 
the fact that Quantum Mechanics is a quantum theory settled on a classical background. If the 
background were quantum as well, the interpretational problem would disappear. Perhaps, the issue of simulations might clarify our point of view, as follows.

\subsection{Simulations}
The issue of simulations of a quantum system has been at the heart of the discovery of quantum 
computers. All that started in 1981, when Feynman \cite{F} proposed for the first time to 
use quantum phenomena to simulate quantum systems.
A classical computer can simulate a quantum system perfectly, as both the computer's memory 
(bits) and the quantum spectra are discrete, but very slowly (in exponential time). Instead, 
a quantum computer can simulate a quantum system perfectly because both the quantum register 
(qubits) and the spectra of the quantum system are discrete, and efficiently, because of 
quantum parallelism \cite{D}.   
However, both the quantum system to be simulated and the quantum computer lie on a classical 
space-time background. The classical background is present before the start of the simulation, 
when a classical input is provided, and at the end, when the observation takes place: at this 
point a large amount of quantum information becomes hidden. This is a kind of inconsistency of 
the whole simulation process, and is due to the fact that Quantum Mechanics is a quantum theory 
formulated on a classical background.
During the very computational process, the classical background is not taken into account, and 
one could guess what is going on inside the ``black box" by performing a new kind of measurement 
which is unitary (reversible) and does not destroy the superposed state. To do so, the observer 
should be internal, that is, she should enter a quantum space whose states are in a one-to-one 
correspondence with the machine states \cite{Z}.

\subsection{The Standard (Projective) Quantum Measurement}
We start by reminding the modalities of the standard quantum measurement (of one qubit). 
Let us consider a qubit in the superposed state:
\begin{equation}
\ket{\psi}=
a\ket{0} + b\ket{1}\label{qubit}
\end{equation}                                                                                                                         

Where $\ket{0}$  and $\ket{1}$  form an orthonormal basis, called the computational basis, and $a$ and $b$, 
called probability amplitudes, are complex numbers such that the probabilities sum up to one:  
$|a|^2 + |b|^2 = 1$.

In vector notation we have: 
$$
\ket{0}\equiv \left(\begin{array}{c}1\\0\end{array}\right)\qquad
\ket{1}\equiv \left(\begin{array}{c}0\\1\end{array}\right).
$$
The standard quantum measurement of the qubit $\ket{\psi}$  in (\ref{qubit}) gives either  
$\ket{0}$ with probability $|a|^2$, or $\ket{1}$  with probability $|b|^2$. 
This is achieved by the use of projector operators. A projector operator $P$ is defined by:
\begin{equation}
P^2=P\qquad
P^+=P
\end{equation}
\noindent
(Where $P^+$ is the Hermitian adjoint of $P$, that is, the conjugate transpose: 
$P^+\equiv P^{T^*}$).
Thus a projector $P$ is idempotent and Hermitian.

Let us consider a general superposed quantum state: $\psi=\sum_{i=1}^n c_i\ket{\psi_i} $
in the Hilbert space $\mathbf{C}^n$, with $\sum_{i=1}^n |c_i|^2 = 1$. The probability $p_r(i)$ 
of finding the state $\psi$  
in one of the basis states $\ket{\psi_i}$ is, after a measurement: $p_r(i)=|P_i\ket{\psi}|^2$. 
After the measurement, the state $\ket{\psi}$ has ``collapsed" to the state 
$\ket{\psi'}=\frac{P_i\ket{\psi}}{\sqrt{p_r(i)}}$.
The $n$ projectors  $P_i$ ($i=1,2,\dots,n$) are orthogonal: $P_iP_j=\delta_{ij}P_i$
and sum up to $1$: 
\begin {equation}
\sum_{i=1}^n P_i=1
\end{equation}
In our case, $i=0,1$. We have the two 2-dimensional projectors:

\begin{displaymath}
P_0=\left(\begin{array}{cc}1 & 0\\0 & 0\end{array}\right)\qquad\\
P_1=\left(\begin{array}{cc}0 & 0\\0 & 1\end{array}\right)
\end{displaymath}
For which it holds:
$$
P_0P_1=P_1P_0=0, \quad P_0^2=P_0, \quad P_1^2=P_1, \quad P_0+P_1=1
$$
\noindent
The actions of $P_0$ and $P_1$ on the basis states are, respectively:
$$
P_0\ket{0}=\ket{0}, \quad P_0\ket{1}=0
\\
P_1\ket{0}=0, \quad P_1\ket{1}=\ket{1}
$$
\noindent
From which it follows that their action on the superposed state (\ref{qubit})
 is, respectively:
 $$
P_0\ket{\psi}=a\ket{0}, \quad P_1\ket{\psi}=b\ket{1}.
$$
\noindent           
The probability of finding the qubit state (\ref{qubit}) in the state $\ket{0}$ is, for example:  
$$
p_r(0)=|P_0\ket{\psi}|^2=|a\ket{0}|^2=|a|^2.
$$
\noindent
After the measurement, the qubit \ref{qubit} has ``collapsed" to the state:
$$
\ket{\psi'}= \frac{P_0\ket{\psi}}{\sqrt{p_r(0)}}=\frac{a\ket{0}}{\sqrt{|a|^2}}=\ket{0}.
$$
Then, a lot of quantum information that was encoded in \ref{qubit} is made hidden by the 
standard quantum measurement. As a projector is not a unitary transformation, the standard 
quantum measurement is not a reversible operation. This means that the hidden quantum 
information will never be recovered (i.e., we will not be able to get back the superposed 
state (\ref{qubit}).

\subsection{The Bloch Sphere}
We believe that the irreversibility of the standard quantum measurement is strictly related 
to the classical geometry of the space-time background. To see why, let us consider the Bloch 
sphere, which is the sphere  with unit radius: 
$$
S^2=\{x_i\in R^3|\sum_{i=1}^3x_i^2=1\}
$$
Any generic 1-qubit state in (\ref{qubit}) can be rewritten as:
 
$$
cos\frac{\theta}{2}\ket{0}+e^{i\phi}sin\frac{\theta}{2}\ket{1}
$$
Where the Euler angles $\theta$  and $\phi$  define a point on the unit sphere $S^2$. 
Thus, any 1-qubit state can be visualized as a point on the Bloch sphere, the two basis states 
being the poles. See Fig.1.
A standard quantum measurement of one qubit is then equivalent to the projection of one of the 
poles of the Bloch sphere, resulting in one point in $R^3$, where the external observer is placed. 
We wish to remind that any transformation on a qubit during a computational process is a 
reversible operation, as it is performed by a unitary operator $U$ such that $U^+U=I$. 
This can be seen geometrically as follows. 
Any unitary $2\times 2$  matrix $U_2$ on $C^2$, (which is an element of the group SU(2) 
multiplied by a global phase factor): 

\begin{equation}\label{fuzzy}
U_2=e^{i\phi }
\left(\begin{array}{cc}\alpha  & \beta \\ 
-\beta^* & \alpha^*\end{array}\right)
\end{equation}                                           	                                                                    
(where $\alpha^*$ is the complex conjugate of $\alpha$ and $|\alpha|^2 + |\beta|^2 =1$), 
can be rewritten in terms of a rotation of the Bloch sphere:
$$
U_2=e^{i\phi }R_{\hat{n}}(\theta)
$$ 
Where  $R_{\hat{n}}(\theta)$ is the rotation matrix of the Bloch sphere by an angle $\theta$  
about an axis $\hat{n}$. 

However, a projector is not a unitary operator, and it cannot be rewritten in terms of a 
rotation 
of the Bloch sphere. This means that the observer, who has performed the standard quantum 
measurement, is not able to recover the original state by a rotation of the sphere. In fact, 
what the external observer sees, is just one pole.

\subsection{The Fuzzy Sphere}
 The question is now whether a reversible measurement could be feasible, at least in principle. 
Of course, the projector should be replaced by a unitary operator, but this means that the 
reversible measurement should be performed ``from inside". Or, in other words, the hypothetical 
observer should be placed in a quantum space whose states are in a one-to-one correspondence 
with the quantum computational states.
 This quantum space will be a discrete topological space associated, by the non-commutative 
version of the Gelfand-Naimark theorem \cite{KW}, with the algebra of 
quantum logic gates. Now, $n$-dimensional quantum logic gates are unitary $n\times n$  complex 
matrices, with $n=2^N$,  where $N$ is the number of qubits in the quantum register. For example,
 in the case of 
one qubit, the quantum logic gates are $2\times 2$   unitary matrices.
Thus, quantum logic gates form a subset of the set of V  complex matrices, whose algebra is a 
non-commutative C*-algebra \cite{Dx}.

To this algebra it is associated, by the non-commutative version of the Gelfand-Naimark theorem,
 a quantum space which is the fuzzy sphere \cite{Mad} with $n$ elementary cells. 
This means that the computational state of a quantum computer with $N$ qubits can be 
geometrically viewed as a fuzzy sphere with $2^N$ cells.

We recall here that the fuzzy sphere is constructed replacing the algebra of polynomials on the 
(unit) sphere $S^2$  by the non-commutative algebra of complex   matrices, which is obtained by 
quantizing the coordinates $x_i$ ($i=1,2,3$): $x_i\rightarrow X_i=kJ_i$, where the $J_i$  
form the $n$-dimensional irreducible 
representation of SU(2): $|J_iJ_j|=i\epsilon_{ijk}J_k$
  and the non-commutative parameter $k$ is, for a unit radius:  
$$k=\frac{1}{\sqrt{n^2-1}}.
$$
Then, the ensemble of rotations of the Bloch sphere (unitary transformations of one qubit) 
can be viewed as a fuzzy sphere in the $n=2$ case (two elementary cells), the $x_i$ being 
replaced by: $x_i\rightarrow X_i= \frac{1}{\sqrt{3}}\sigma_i$, where the $\sigma_i$  are the 
Pauli matrices.

\subsection{The Mirror Measurement}
In what follows, we will generalize the standard quantum measurement of one qubit by using   
complex matrices that are not projectors, and we will analyze the associated geometries.

To start, let us consider the diagonal   matrices on the complex numbers (they form a 
commutative C*-algebra, which is a sub algebra of the non-commutative C*-algebra of $2\times 2$
matrices on the complex field). Recall, however, that we shall require unitarity, so that we 
should consider only matrices of the kind: 
\begin{equation}\label{mirror}
U_2=e^{i\phi }
\left(\begin{array}{cc}\alpha  & 0 \\ 
0 & \alpha^*\end{array}\right)
\end{equation}    
With $|\alpha|^2=1$. Where $U_2^D$  in (\ref{mirror}) is the particular case of  $U_2$  in 
(\ref{fuzzy}), with $\beta =0$.
The action of $U_2^D$  on the qubit state (\ref{qubit})  gives: 
$$
U_2^D\ket{\psi}=a'\ket{0}+b'\ket{1}
$$                                                                                                                
Which is still a superposed state with: $a'=e^{i\phi}\alpha a, b'=e^{i\phi}\alpha^*b$, and:
$$
|a'|^2=|a|^2\\qquad|b'|^2=|b|^2.
$$  
That is, the probabilities are unchanged. Notice that geometrically, this is equivalent to 
project both the poles of the Bloch sphere at the same time. The associated space is a 
2-points lattice (which is a discrete, but still classical, space).
In fact, $U_2^D$  can be rewritten as:
\begin{equation}\label{sovrapp}
U_2^D=e^{i\phi}(\alpha P_0 + \alpha^* P_1)
\end{equation}                                                                                                             
Which is a linear superposition of the two projectors $P_0$ and $P_1$. Then, $U_2^D$
  is the reversible origin of a standard (irreversible) quantum measurement. The application 
of $U_2^D$    to the state  $\ket{\psi}$ in (\ref{qubit}) is a superposition of two standard 
quantum measurements made simultaneously. We will call this new kind of quantum measurement 
{\em ''mirror-measurement"} because the qubit remains in a superposed state, and the probabilities are 
unchanged.

After the mirror-measurement, the state  $\ket{\psi}$ is left in the state:
\begin{equation}\label{}
\ket{\psi }\rightarrow \ket{\psi'}=\frac{U_2^D\ket{\psi }}{\sqrt{|a'|^2+|b'|^2}}=
e^{i\phi} (\alpha a\ket{0}+\alpha^*b\ket{1}
\end{equation}
Where one has considered the total probability. 

The state $\ket{\psi'}$ is still a superposed state, and, from it, one can recover the original state   
by performing the inverse operation:
\begin{equation}\label{}
(U_2^D)^{-1}\ket{\psi'}
=e^{-i\phi} \left(\begin{array}{cc}\alpha^*  & 0\\ 
0 & \alpha\end{array}\right)
\left(\begin{array}{c}e^{-i\phi}\alpha a\\ 
e^{-i\phi}\alpha^*b\end{array}\right)=
\left(\begin{array}{c} a\\ b\end{array}\right)=\ket{\psi }
\end{equation} 
In summary, the mirror measurement does not destroy the superposition, since it is reversible, 
and does not change the probabilities, but just changes the probability amplitudes.

It should be noticed that the internal observer (who will be called {\bf P}
 in the following) uses the projectors $P_0$ and $P_1$  and   at the same time. She can do so 
as she lives in a discrete space, namely a 2-points lattice, which is in a one-to-one 
correspondence with the two basis states, as it is the space associated with the algebra of 
diagonal unitary $2\times 2$  matrices $U_2^D$. 
If an external observer ({\bf G}) should try to do the same, she would 
fail, as she lives in a classical, continuous space, namely, $R^3$. Or, {\bf G} could try to 
achieve the 
same result of {\bf P} by using  $P_0$ and $P_1$ in parallel on two copies of the same state 
$\ket{\psi }$, but that is 
forbidden by the no-cloning theorem \cite{WZ}, which states that an unknown 
quantum state cannot be copied. Then, the only thing that {\bf G} can do, is to use either $P_0$
or  $P_1$, 
that is, to perform a standard quantum measurement. The action of {\bf G} then breaks the 
superposition of  $P_0$ and $P_1$, used by {\bf P}. 

Finally, it can be seen that the mirror measurement performed by {\bf P} is not in contradiction 
with the standard quantum measurement made by {\bf G}. 
In fact, if {\bf G} uses $P_0$ on the state $\ket{\psi'}= U_2^D\ket{\psi }$, 
(that is, she performs a standard measurement after a 
mirror measurement), she gets:
\begin{equation}\label{}
P_0U_2^D\ket{\psi }=P_0[e^{i\phi}(\alpha P_0 + \alpha^*P_1)](\ket{\psi'}= a'\ket{0} 
\end{equation}                                                                          
With probability  $|a'|^2=|a|^2$.
After this ``composed" measurement, the qubit is left in the state:
\begin{equation}\label{} 
\ket{\psi }'=\frac{P_0U_2^D\ket{\psi} }{\sqrt{p_r(0)} }=\frac{a'\ket{0} } {\sqrt{|a'|^2}}=\ket{0}   
 \end{equation}                                             
In other words, the result obtained by {\bf G} in the classical world is not influenced by the result obtained by 
{\bf P} in the quantum world. {\bf P}'s operation gives a result that is consistent with the standard quantum 
measurement. Or, in other words, she does not create contradiction to the external observer. This is very important 
not only in the physical sense, but also in the logical sense, as we will see in the next sections.
In passing from the mirror measurement to the standard quantum measurement, the associated geometry has changed: 
from the 2-points lattice (the two poles of the Bloch sphere) to one point (one pole of the Bloch sphere). 
In the context of the mirror measurement, it should be noticed, however, that the 2-points lattice breaks 
rotational invariance, so that, from this space, {\bf P} cannot reach any other 1-qubit state of the Bloch sphere, 
by a rotation. She can just make a quite limited operation that, up to a global phase factor, is just a phase shift,
 as showed below.
 
Any 1-qubit unitary transformation $U_2$ can be written as:
$$
U_2=e^{i\phi}R_Z(\gamma)R_\gamma(\theta)R_Z(\delta)
$$
with:
\begin{displaymath} 
R_Y(\theta)=e^{-i\theta Y/2}=cos\frac{\theta}{2}I-i sin\frac{\theta}{2}Y=
\left(\begin{array}{cc}cos\frac{\theta}{2}  & -sin\frac{\theta}{2} \\ 
sin\frac{\theta}{2} & cos\frac{\theta}{2} \end{array}\right)
\end{displaymath} 
\begin{displaymath} 
R_Z(\delta)=e^{-i\delta Z/2}=cos\frac{\delta}{2}I-i sin\frac{\delta}{2}Z=
\left(\begin{array}{cc}e^{-i\delta /2}  &  0\\ 
0  & e^{i\delta /2}\end{array}\right)
\end{displaymath} 
Where $Y$ and $Z$ are the Pauli matrices:
\begin{displaymath} 
Y=
\left(\begin{array}{cc}0 & -i\\ 
i & 0 \end{array}\right)
\qquad
Z=
\left(\begin{array}{cc} 1 &  0\\ 
0  & -1 \end{array}\right)
\end{displaymath} 
With the choice $\theta=0$ and $\gamma=\delta$ one gets:
\begin{equation} \label{expmirror}
U_2=e^{i\phi} 
\left(\begin{array}{cc}e^{-i\delta /2}  &  0\\ 
0  & e^{i\delta /2}\end{array}\right)
\end{equation} 
Which is our $U_2^D$   matrix in (\ref{mirror})
with $\alpha=e^{-i\delta}$
(recall that $\alpha$  is a complex number with unit modulus).
Finally, the diagonal matrix in (\ref{expmirror})
can be written as:
\begin{equation} \label{expmirror2}
U_2=e^{i\phi'} 
\left(\begin{array}{cc} 1 &  0\\ 
0  & e^{i\lambda}\end{array}\right)
\end{equation} 
With: $\phi'=\phi+\delta$, and $\lambda=2\delta$ . The matrix in (\ref{expmirror2})
is, up to the global phase factor 
$e^{i\phi}$, the quantum gate ``phase shift".

By Eq. (\ref{expmirror2}), then, the mirror measurement corresponds to an anti-clockwise rotation about the 
z-axis of the Bloch sphere. See Fig. 2.

This is equivalent to the fact that {\bf P}
 stands on a 2-points lattice embedded in the fuzzy sphere.
 
On the other side, Eq. (\ref{sovrapp})
provides the geometrical interpretation of the mirror measurement as conceived by {\bf G}, 
from outside: an ordered 2-points lattice, without the embedding in the fuzzy sphere.
Of course, it is also possible to perform a mirror measurement in a different basis, for example, in the dual basis,
$\ket{\pm }=\frac{1}{\sqrt{2}}(\ket{0}+\ket{1})$,        
which is obtained by applying the Hadamard gate: $H=\frac{1}{\sqrt{2}}
\left(\begin{array}{cc} 1 &  1\\ 
1  &  -1\end{array}\right)$
   to the computational basis states $\ket{0}$ and $\ket{1}$ respectively. 
In the dual basis, the two orthogonal projectors are: 
$$
P^+=HP_0H^{-1},\qquad P^-=HP_1H^{-1}
.
$$
 
The diagonal unitary operator in (\ref{mirror}) is transformed as: 
$$
U_2^D \rightarrow  (U_2^{D})'=H U_2^D H^{-1}
$$
Which can be written as a linear superposition of $P^+$ and $P^-$:
$$
U_2^D=e^{i\phi}(\alpha P^+ + \alpha^* P^-)
$$ 
That is, the mirror measurement in the dual basis can be viewed as the simultaneous actions of two orthogonal 
projective measurements in that basis.

\subsection{The Fuzzy Measurement}
If {\bf P} wished to reach any other point of the Bloch sphere, she should not limit herself to the diagonal 
$2\times 2$  
unitary matrices, but should consider the full algebra of  $2\times 2$ unitary matrices, which is a non-commutative 
C*-algebra. 

Notice that the 2-point lattice considered above, is a sub-space \cite{Mar} of the fuzzy sphere \cite{Mad}. 
When instead of considering the unitary diagonal $2\times 2$ matrices as in (\ref{mirror}), one considers the full 
algebra of 
unitary  $2\times 2$ matrices, the two points of the lattice are ``smeared out" into two cells of a fuzzy sphere. 
See Fig.3. 
Then, the original qubit has not just been ``phase shifted" but has been rotated, so that its original probability 
amplitudes have been ``mixed up".
In fact, the action of a unitary  $2\times 2$ matrix on the qubit state (\ref{qubit}) is:
\begin{equation} \label{}
U_2\ket{\phi}=e^{i\phi} 
\left(\begin{array}{cc} \alpha &  \beta\\ 
-\beta^*  & \alpha^*\end{array}\right)
\left(\begin{array}{c} a\\ 
b \end{array}\right)=
e^{i\phi} 
\left(\begin{array}{c} \alpha a + \beta b\\ 
-\beta^* a + \alpha^*b \end{array}\right)=
\ket{\phi}'
\end{equation} 
                                               
To summarize, {\bf P}  should place herself in a fuzzy sphere if she wishes to follow the whole computational 
process from inside the quantum computer, but she can just stand in a subspace of the fuzzy sphere, namely the 
2-points lattice, if she wants to perform a mirror measurement. Notice that when {\bf P}  is performing a generic   
$L_2$-operation on the qubit, she has lost any contact with the external world, and cannot communicate anymore 
with {\bf G}, 
as now {\bf G}  can only interpret the world of {\bf P}  as a ``black box". This will be shown in logical terms 
in the next 
sections.

\subsection{The Liar Measurement}
We finally introduce a particular kind of reversible measurement, that we will call 
``Liar measurement". Algebraically, it consists of an off-diagonal unitary matrix:
\begin{equation}\label{liar}
L=e^{i\phi }
\left(\begin{array}{cc}0 & \alpha \\ 
\alpha^* & 0\end{array}\right)
\end{equation}                                                                                                      
Which is obtained by applying a NOT gate after a diagonal unitary matrix:
$$
L={\mbox NOT}\,U_2^D
$$  
Geometrically, it corresponds to a clockwise rotation around the z-axis of the Bloch sphere, where the North pole 
is $\ket{0}$  and the South pole is  $\ket{1}$, as usual. As it is well known, a clockwise rotation is equivalent 
to an anti-clockwise
 rotation with the poles interchanged. See Fig.4. Both these geometrical pictures are relative to the internal 
observer {\bf P} when she stands on the 2-points lattice inside the fuzzy sphere.
Moreover, since from Eq. \ref{sovrapp}, we know that $U_2^D$ can be written as:
$$
U_2^D=e^{i\phi}(\alpha P_0 + \alpha^* P_1),
$$ 
we get: 
\begin{equation}\label{liarsovrapp}
L=e^{i\phi}(\alpha Q_0 + \alpha ^* Q_1)
\end{equation}                                                                                                               
Where: 
\begin{displaymath}\label{}
Q_0={\mbox NOT} P_0=\left(\begin{array}{cc}0 & 0 \\ 
1 & 0\end{array}\right)
\qquad
Q_1={\mbox NOT} P_1=\left(\begin{array}{cc}0 & 1 \\ 
0 & 0\end{array}\right)
\end{displaymath} 
Obviously, $Q_0$  and $Q_1$  exchange the truth-values:
$$
Q_0\ket{0}=\ket{1} \qquad Q_0\ket{1}=0 
\\
Q_1\ket{0}=0 \qquad Q_1\ket{1}=\ket{0}
$$
Eq. (\ref{liarsovrapp}) gives the interpretation of the Liar measurement by the external observer who conceives the 
ordered 2-points lattice with the two elements interchanged without the embedding in the fuzzy sphere. 
The action of $L$ on the qubit $\ket{\psi}$  is:
\begin{equation}\label{liarappl}
L\ket{\psi}=a'\ket{0} + b'\ket{1}
\end{equation}                                                                                                                       
where: $a'=e^{i\phi}\alpha^*b$  and  $b'=e^{i\phi}\alpha a$
Then, the two probabilities are interchanged.
If the external observer {\bf G} measures  $\ket{\psi}$ after {\bf P} has performed a Liar measurement, for example, 
by means of the projector $P_0$:
\begin{displaymath}\label{}
P_0 L\ket{\psi}=a'\ket{0}
\end{displaymath}                                                                                                           
She gets $\ket{0}$ with probability $|b|^2$.

\section{Logical Interpretation of the Mirror and Liar Measurements}\label{logica}
In this paper, we have defined ``black box" the computational state of a quantum computer, which cannot be known by 
an external observer {\bf G}. In fact, {\bf G} can only achieve a small part of the quantum information being 
processed, when she 
performs a standard quantum measurement. 
The aim of this section is to provide a logical interpretation of quantum computation, from the point of view of an 
observer {\bf P} who is inside the black box. Or, which is the same, {\bf G} is in a quantum space like the fuzzy 
sphere, described in the previous section, whose states are in a one-to-one correspondence with the machine states.

\subsection{How the Insider Observer Gets Rid of the Non-Contradiction Principle }
We confine ourselves to the simplest case, a toy model-quantum computer based on a quantum register of one qubit. 
The qubit is a vector of ${\mathbf C}^2$. Let us consider an orthonormal basis of  ${\mathbf C}^2$, that is a couple 
of orthonormal states $A, A^\perp$. The state of the qubit is a superposition of the two basis states: we will see 
now the effect of performing a measurement on it in logical terms. 
We first consider a standard quantum measurement, performed by the external observer {\bf G}. For example, let us 
suppose that the projector $P_0$  with respect to the first element of the orthonormal basis is applied, obtaining 
the state $A$. Then the external observer puts the judgement $\vdash A$,  meaning: ``the qubit is in the state $A$"
 \cite{B}. Analogously, by the application of the projector $P_1$ , the observer would obtain the state 
$A^\perp$,
 and then she would put the judgement $\vdash A^\perp$.
 
Now, we consider the reversible measurement discussed in Sect. \ref{misure}. Eq. (\ref{sovrapp}) means that the 
external observer can 
interpret the mirror measurement as the superposition of two orthogonal projectors. Then, in logical terms, the 
reversible measurement gives back a {\em couple} of judgements: 
\begin{equation}\label{coppiavero}
\vdash A \qquad\quad \vdash A^\perp
\end{equation}
Which are given by the simultaneous actions of  $P_0$ and $P_1$. 
Now, as in \cite{B}, we follow the reflection principle in \cite{SBF}. By the reflection principle, 
the logical connectives are the result of importing some existent meta-linguistic links between assertions into the 
formal language. Let us consider the physical link of superposition between orthogonal states, which is possible only
 inside the quantum computer. This becomes a logical link between opposite judgements once the superposition has been
 measured, obtaining the couple of judgements above.
We interpret the juxtaposition of two judgements by the additive conjunction $\&$ and put:
\begin{equation}\label{eqvero}
\vdash A\& A^\perp \quad\equiv\quad \vdash A \qquad \vdash A^\perp
\end{equation}
The external observer interprets the reversible measurement by means of (\ref{eqvero}), or by the inference 
derived from it:
\begin{equation}\label{dervero}
\frac{\vdash A \quad \vdash A^\perp}{\vdash A\& A^\perp}
\end{equation}                                                                                                                                                     
We remind that  the internal observer is unaware of the possibility of writing $U_2^D$  as a linear superposition of 
projectors, as to her, projectors are meaningless, being non-unitary, thus not belonging to the quantum network. 
Then,  the couple of judgements (\ref{coppiavero}) is not available to her, so that she cannot perform the derivation 
(\ref{dervero}): she  just interprets the measurement of the superposed state as the axiom:
\begin{equation}\label{axvero}
\vdash A\& A^\perp
\end{equation}
Which is the opposite of the non-contradiction principle.  
                                         
Axiom (\ref{axvero}) coincides with the conclusion of the external observer in (\ref{dervero}). 
To the internal observer, 
the axiom (\ref{axvero}) means precisely the following: ``The superposed state has been measured by a mirror 
measurement". 
Then, the proposition ``$\vdash A\& A^\perp$", represents, in logical terms, the superposition of the two orthogonal 
states, in a qualitative way, without taking into account the probability amplitudes.
As a diagonal quantum logic gate, which does not change the truth-values, performs the mirror measurement, 
we say that a superposed state is ``measurable" if and only if the following axiom holds:
\begin{equation}\label{axseq}
A \& A^\perp\vdash A\& A^\perp
\end{equation}                                                                                                                  
Where here the sequent symbol $\vdash$ must be read: the transition is done by means of a $U_2^D$ 
(a generalized sequent calculus for quantum computing, comprising non diagonal unitary transformations, 
is under study \cite{BZ3}.

\subsection{How the Internal Observer Gets Rid of the Excluded Middle} 
Let us suppose now that the external observer applies a standard quantum measurement to the qubit and then decides 
to apply a classical NOT gate on the result.
If she had obtained, for example  after the measurement, she would get $A^\perp$ 
 after the NOT, but now she cannot assert  $A^\perp$ as true, since it is not the result 
of her measurement! In fact, the correct assertion is that $A^\perp$  is false. 
We write such assertion as in \cite{SBF}: $A^\perp\vdash$, that is a primitive way to say that $A^\perp$  
is false. The composite 
operation just discussed, converted the original judgement $\vdash A$  into $A^\perp\vdash$.  So, we put the 
equivalence:
$$(\vdash A)^\perp\equiv A^\perp\vdash  .
$$
If instead she measured  $A^\perp$ , and performed a NOT afterwards, she would get the judgement: $A\vdash$ .
If by absurd, she could perform the two measurements simultaneously, and apply a NOT afterwards, she would obtain 
both the judgements:
$$
A^\perp\vdash \qquad \quad A\vdash
$$                                                                                                                                                        
This is a logical link between two ``falsity judgements", and, as in \cite{SBF}, it is solved as:
\begin{equation}\label{eqfalso}
A\oplus A^\perp\vdash \quad\equiv\quad A \vdash\qquad\quad A^\perp\vdash
\end{equation}                                                                                                        
 Where $\oplus$  is the additive logical disjunction. Let us recall, however, that the external observer {\bf G} 
cannot perform the two measurements simultaneously, while the internal observer {\bf P}  can. Then, {\bf P}  puts 
the following falsity judgement that is an axiom:
 \begin{equation}\label{axfalso}
A^\perp\oplus A\vdash  
\end{equation}                                                                                                                
which implies that the excluded middle principle does not hold inside the quantum computer. Axiom (\ref{axfalso}) 
states that 
the superposed state has been measured by a Liar measurement.
In logical terms, the Liar measurement means that inside the quantum computer the judgement $\vdash A \& A^\perp$   
has been reversed: $A^\perp \oplus A\vdash$.
Axiom (\ref{axfalso}) is so recovered from axiom (\ref{axvero}) by symmetry, a fundamental feature of Basic Logic 
\cite{SBF}. 
The connective $\oplus$ is the symmetric of $\&$ \cite{SBF} since propositions ''$A^\perp\oplus A$ and  
``$A\& A^\perp$" both represent the 
superposed state. In conclusion, $\oplus$  and $\&$ are logical connectives corresponding to the same meta-linguistic
 link: superposition.
 
The axiom:
\begin{equation}\label{axseqfalso}
A^\perp\oplus A\vdash  A^\perp\oplus A
\end{equation}                                                                                                                        
Is obtained by making the symmetric of axiom (\ref{axseq}), and must be interpreted in the following way: a 
superposed state, which has already been measured by a Liar measurement, can be measured by a mirror measurement 
performed by  $U_2^D$. 
Axioms (\ref{axvero}) and (\ref{axfalso}) state that, in the black box, both the non-contradiction and the excluded middle
 principles do not hold. This means that inside the black box the adequate logic is a paraconsistent and symmetric 
logic, like Basic Logic.
The additive connectives $\&$ and $\oplus$  are the only ones we can consider, when we deal with a 1-qubit model. 
\subsection {The Qubit and the Mirror: Some Philosophical Implications}
To us, the philosophical meaning of axiom (\ref{axseq} is the following. It looks like the superposed state reflects 
itself 
in a slightly deformed mirror, that is, the diagonal unitary operator  , which just changes the probability 
amplitudes, but leaves unchanged the truth-values (the identity operator being the perfect mirror). This analogy 
would suggest that the qubit has gone through a self-measurement. The act of  ``looking at itself in the mirror " 
confirms the existence of the qubit as it stands. This is what we would call {\em objectivity} of an elementary 
quantum 
system. This is on the same line of thought followed by Mermin in \cite{Mer}. Among his six desiderata for an 
interpretation of quantum mechanics, the first concerns objective reality which should be separated from external 
observers and their ``knowledge", and this is in fact our case, as we do not add external judgements to the logical 
interpretation of the insider observer. 
The fourth desideratum is that quantum mechanics should not require the existence of a classical domain, which is in 
fact our case, as the internal observer just represents a quantum mechanical system in a quantum domain (the fuzzy 
sphere). The sixth desideratum, that the probabilities should be objective properties of individual systems, is 
fulfilled in the mirroring of the qubit.  
We think that this philosophical interpretation might be useful in the case of two entangled qubits, as the ``spooky 
action at a distance", as seen by an external observer might be related to a reversible measurement of the entangled 
state.

\section{The Border between the Black Box and the Classical World}\label{border}
In Sect. \ref{logica}, we gave a logical description of the reversible quantum measurement performed 
in the black box. To be exhaustive, however, we should also understand the right way to pass through the border 
between the black box and the classical world, (the infamous measurement problem) in logical terms. To this aim, we 
will exploit axioms \ref{axvero} and \ref{axfalso}. To illustrate the border of the black box, we present two 
possible schemes: the first is physical, the second is logical.

\subsection{The Physical Scheme}
The physical scheme is the following: a classical input is provided to the quantum computer from the classical world;
 the quantum computation is the black box. In the black box, an insider observer P performs a reversible quantum 
measurement.
A classical output is obtained by the action of an external observer {\bf G}, who, from outside, opens the quantum 
system by performing a standard quantum measurement. In the physical scheme, then, the border coincides with the 
standard (projective) quantum measurement process. See Fig. 5.

\subsection{The Logical Scheme}
The logical scheme is the following: if the black box is to be interpreted by an external logician, 
it can be considered embedded in the classical world at the left and right sides (past and future).
On the left side, from where the classical input is provided, we can imagine an external observer 
{\bf A} (Aristotle) who can reason by classical logic. In the black box, we have a new kind of quantum 
logic which is the logic of the insider observer {\bf P}. She can manage the two new axioms \ref{axvero}
and \ref{axfalso} as far as the border. On the right side of the black box, however, one cannot immediately
place {\bf A} reading the classical output. As, in that case, {\bf A} would receive the two new axioms
that for him, who is a classical logician, are contradictions. In fact, {\bf A} can identify 
$A^\perp$
as follows:
\begin{equation}\label{falso}
A^\perp\equiv A\rightarrow \perp
\end{equation}
Also, he can use the {\em modus ponens} rule:
\begin{equation}\label{mp}
\frac{A  \quad A\rightarrow \perp}{\perp}
\end{equation}
The identity \ref{falso} and the rule\ref{mp} would then allow to derive the {\em falsum} from the premises
$A$ and $A^\perp$.

Therefore, prior to {\bf A}, there must be another external observer, who is a quantum logician: {\bf G}. In fact,
{\bf G} cannot apply \ref{falso} and \ref{mp}, as her logic is too weak. She is able to drop the two new axioms
in the following way: she applies the cut rule to \ref{axfalso}, obtaining the following derivation, valid in quantum 
logic:
\begin{equation}\label{cut}
\frac{\vdash A\& A^\perp  \quad \frac{A\vdash A}{A\& A^\perp\vdash A}\,\& L
}{\vdash A}\,cut
\end{equation}
Then, from \ref{cut} one can see that the cut rule, when applied to the axiom \ref{axvero}, is equivalent to the 
physical demolition of the superposed state, as it ''creates a projector".

It should be noticed that the insider observer {\bf P} can provide one new axiom \ref{axvero} only once, because of
the no-cloning theorem \ref{??}, which forbids the copy of an arbitrary quantum state. This is a very useful theorem, 
otherwise {\bf G} would not be able to drop the axioms (as she does using \ref{cut}), which will then provided to
{\bf A}, as a contradiction!
In fact, let us suppose, by absurd, that the no-cloning theorem did not hold, and {\bf P} could give the axiom
\ref{axvero} to {\bf G} twice. Then {\bf G} could derive first $\vdash A$ and then $\vdash A^\perp$, and finally 
she would derive again $\vdash A \& A^\perp$, as follows: she would perform the derivation \ref{cut} by using the 
axiom \ref{axvero} for the first time. Then, she would perform the same derivation for the second time, by replacing
the sequent calculus axiom $A\vdash A$ by $A^\perp\vdash A^\perp$ and obtaining then $\vdash A^\perp$. At this point,
{\bf G} could apply (\ref{dervero}) and thus get the conclusion $\vdash A\& A^\perp$. 
That is, she would get again the 
axiom (\ref{axvero}) (which is false in the external world) just by logical reasoning, not by a mirror measurement.

However, {\bf G}, having measured $A$ in the standard way, can logically conclude $\vdash A\oplus A^\perp$, and
$A\& A^\perp\vdash$. They are instances of the excluded middle and of the non-contradiction principles. In this case
such instances are like empirical laws, as they follow from a measurement, not from physical reasoning. So, {\bf G}
has a logical point of view opposite to that of {\bf P}.
Notice finally that, if the no-cloning theorem did not hold, {\bf G}, besides providing 
contradiction to {\bf A} as shown above, would also fall in contradiction herself. In fact, 
she could derive axioms \ref{axvero}
 and \ref{axfalso}
 which are the opposite of her results, that is 
non-contradiction and excluded middle. 
In summary, the physical border of the black box corresponds, in logical terms, to the site of 
an intermediate logic (between black box logic and classical logic), which is standard quantum 
logic. See Fig. 6.				
To conclude this section, we wish to illustrate the main differences among observers {\bf A}, 
{\bf G}, 
and {\bf P}, in both the physical and logical ways. Let us consider first the differences in 
the physical sense.
{\bf A} is an external observer, who does not perform any kind of quantum measurement. He lives 
in a classical world, without any interaction with the quantum world. He just provides the 
classical input to the quantum computer, and considers the classical output as an element of 
the classical world (for example one bit of a classical computer) without any link to the 
quantum computer.
{\bf P} is an insider observer: she is in a quantum space whose states are in a one-to-one 
correspondence with the machine states, so that she can perform a reversible measurement, 
without hidden information. 
 {\bf G} is also an external observer like {\bf A}, but she can perform a standard quantum 
measurement. However, {\bf PG} lives in a classical space-time, so that she breaks the 
isomorphism (created by {\bf P}) between observer and machine. For this reason, {\bf G} can 
only perform a standard quantum measurement (which is irreversible). She opens the black box, 
and destroys the superposed state, that is, she is responsible of the hidden quantum information.
Now, let us consider the differences in the logical sense. 
{\bf A} (Aristotle) is a classical logician, who believes in the excluded middle and in the 
non-contradiction principles, only by logical reasoning, in a formal way, as he never does any 
kind of quantum measurement. He is in total contradiction with {\bf P}, so he can never 
communicate 
with her. However, {\bf A} can communicate with {\bf G}, because they are not in contradiction; 
simply, 
{\bf G} has a weaker logic than {\bf A}.
{\bf G} can communicate with both {\bf A} and {\bf P}. {\bf G} believes in the excluded middle 
and non-contradiction, but not in the formal way like {\bf A}, as she relies her beliefs on 
standard quantum measurement. 
The logic of {\bf P} is still under study (see \cite{BZ} for the two-qubits model, in particular the Bell's states); 
at present we only know it is 
paraconsistent and symmetric when {\bf P} stands on a 1-dimensional subspace of the fuzzy sphere. 
However, we foresee the following. Having at her disposal such a big amount of quantum information 
(encoded in a couple of very strong axioms), P will need fewer structural rules than G.
Finally, as we have seen, {\bf P} can communicate with G because of the no-cloning theorem, 
but {\bf P} cannot, by no means, communicate directly with {\bf A}.
In meta-language, let us introduce the new meta-connective @ (communicate). For the previous 
arguments, we have: {\bf A} @ {\bf G}, {\bf G} @ {\bf P} but not {\bf A} @ {\bf P}. That is, 
the transitive rule does not hold for this communication process. 

\section{Conclusions}
In this paper, we have considered quantum computation from the point of view of three different 
observers: {\bf P}, the insider one, and {\bf G} and {\bf A}, the external ones. {\bf P} can 
reason in terms of a new 
quantum logic, which is paraconsistent and symmetric, {\bf G} in terms of standard quantum logic, 
and {\bf A} in terms of classical logic. What made possible this logical distinction is the physical 
distinction between two different kinds of quantum measurement: the reversible one, from inside,
 and the irreversible one, from outside. The physical distinction, moreover, is based on the 
geometrical distinction between quantum and classical backgrounds, which is based itself on the 
distinction between non commutative and commutative C*-algebras. In this way, we put together 
quantum computing (and quantum physics), geometry, algebra and logic, as summarized in Table I.
We are aware of the fact that our toy model is, up to now, limited to the case of only one qubit and two qubits,
for the Bell's states \cite{BZ}. 
To conclude, we would like to make the following remark: we believe that our logical approach to quantum computing might be 
useful in conceiving quantum control in quantum information theory. 
\bigbreak
{\bf Acknowledgements:} 
Work supported by the research project ``Logical Tools for Quantum Information Theory", 
Department of Pure and Applied Mathematics, University of  Padova.

\begin{table} \label{tabella}
\center{{\bf{Table 1} }}
\begin{center}
\begin{tabular}{|l |l |l |l |l |}
\hline
\bf{Physics} & \bf{Geometry} & \bf{Symmetry} & \bf{Algebra} & \bf{Logic}\\
\hline
{\bf Black Box} & {\bf Quantum}  & {\bf Rotational}  & {\bf Non commutat.} & 
{\bf Black Box}\\
Fuzzy & {\bf geometry} &  {\bf invariance}& {\bf $C^\ast$-algebra} & {\bf Logic}\\
measurement & Fuzzy sphere &  & Algebra of & \\
 & with 2 cells &  & unitary $2\times2$ & {\bf under study}\\
  &  &  & matrices on ${\bf C}$ & \\
\hline
\bf{Insider} & \bf{Classical} & \bf{Breaking of} & \bf{Commutative} & \bf{Paraconsistent}\\
\bf{Observer P} & \bf{discrete} & \bf{rotational} & \bf{$C^*$-algebra} & \bf{Symmetric}\\
             & \bf{geometry} & \bf{invariance} &          & \bf{Logic}\\
Mirror & Two-points &        & Algebra of &       \\ 
measurement & lattice &        & diagonal unitary &      \\
No hidden & Subspace of &        & $2\times 2$ matrices &      \\ 
quantum & the fuzzy &        & on $C$    &        \\  
information & sphere &        &         &        \\        
\hline
\bf{Interface} & \bf{One point} & \bf{Breaking of} & \bf{Algebra of} & \bf{Quantum Logic}\\
External & One pole of & \bf{rotational} & \bf{projectors} &            \\
observer \bf{G} & $S^2$ & \bf{invariance} &         &        \\
Projective &   &  &   &  \\
quantum &   &   &   &  \\
measurement. &   &   &   &  \\
Hidden quantum &   &   &   &  \\
information &   &   &   &  \\
\hline
\bf{Outside} & \bf{Classical} & \bf{Rotational} & \bf{Algebra of} & \bf{Classical Logic}\\ 
Classical & \bf{continuous} & \bf{invariance} & \bf{functions on $S^2$}        &        \\
input/output & \bf{geometry} &      &       &       \\
External & The classical  &   &   &  \\
observer {\bf A} & sphere $S^2$  &   &   &  \\
\hline
\end{tabular}
\end{center}
\end{table}

 
\newpage

\begin{figure}{{\large
{\bf
Fig. 1}}}
\medbreak
{\large  The Bloch Sphere}
\label{figura3}
\centering
\resizebox{12cm}{!}
{
\includegraphics*{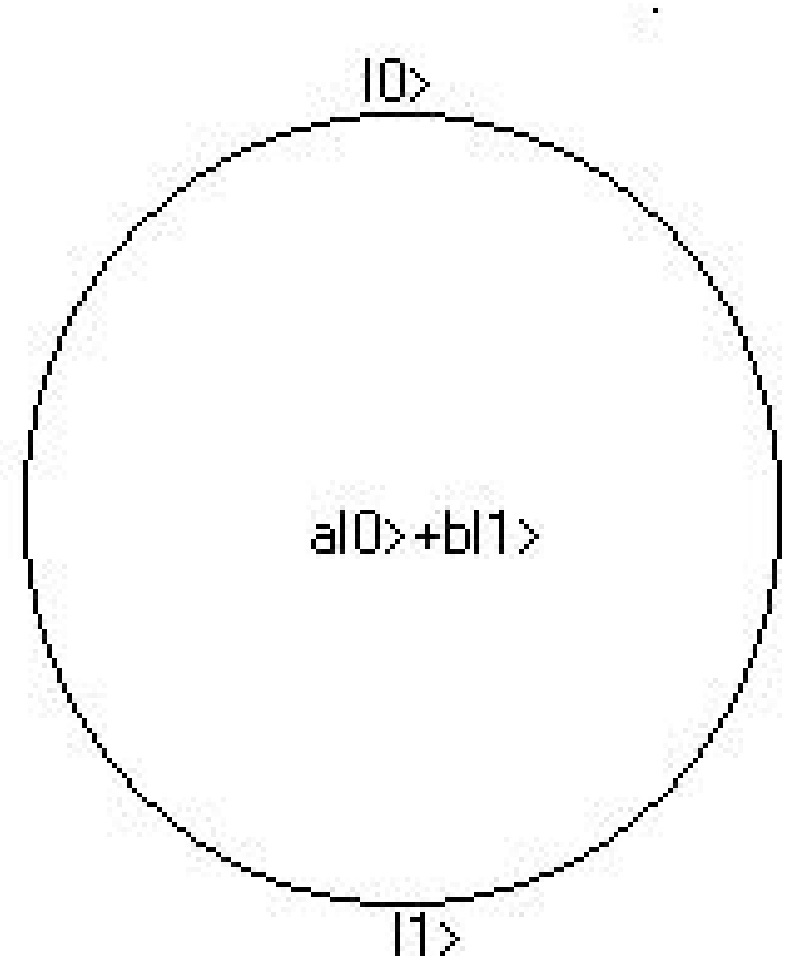}}
\end{figure}
\newpage
\begin{figure}{{\large
{\bf
Fig. 2}}}
\medbreak
{\large  The Mirror Measurement}
\label{figura2}
\centering
\resizebox{12cm}{!}
{
\includegraphics*{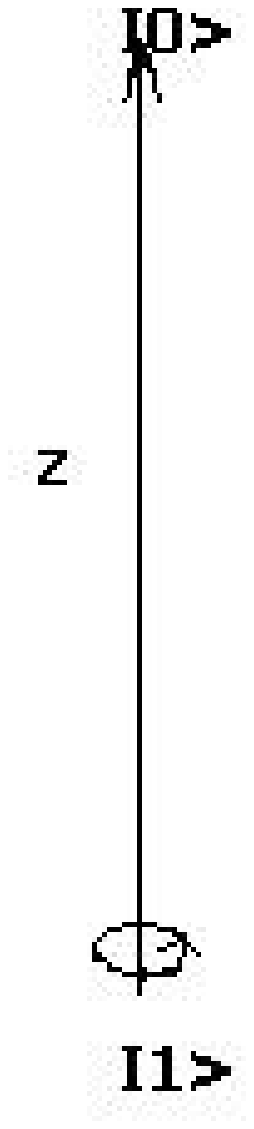}}
\end{figure}
\newpage
\begin{figure}{{\large
{\bf
Fig. 3}}}
\medbreak
{\large  The Fuzzy Measurement}
\label{figura2}
\centering
\resizebox{12cm}{!}
{
\includegraphics*{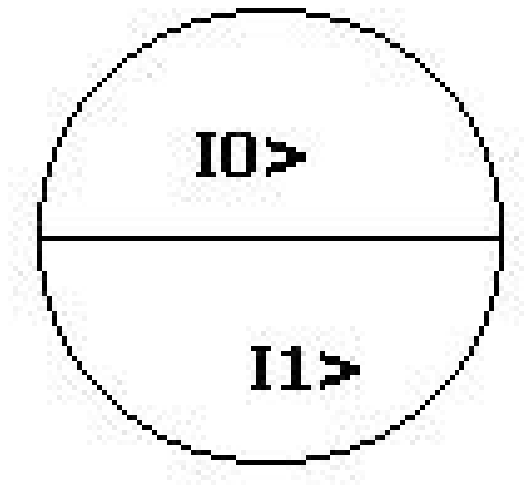}}
\end{figure}
\newpage
\begin{figure}{{\large
{\bf
Fig. 4}}}
\medbreak
{\large  The Liar Measurement}
\label{figura3}
\centering
\resizebox{12cm}{!}
{
\includegraphics*{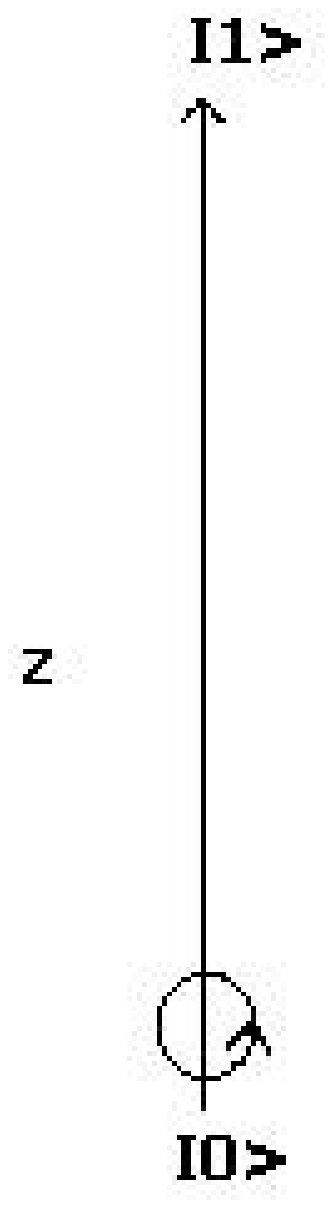}}
\end{figure}
\newpage
\begin{figure}{{\large
{\bf
Fig. 5}}
\medbreak
{\large 
The Physical Scheme}}
\label{figura1}
\centering
\resizebox{12cm}{!}
{
\includegraphics*{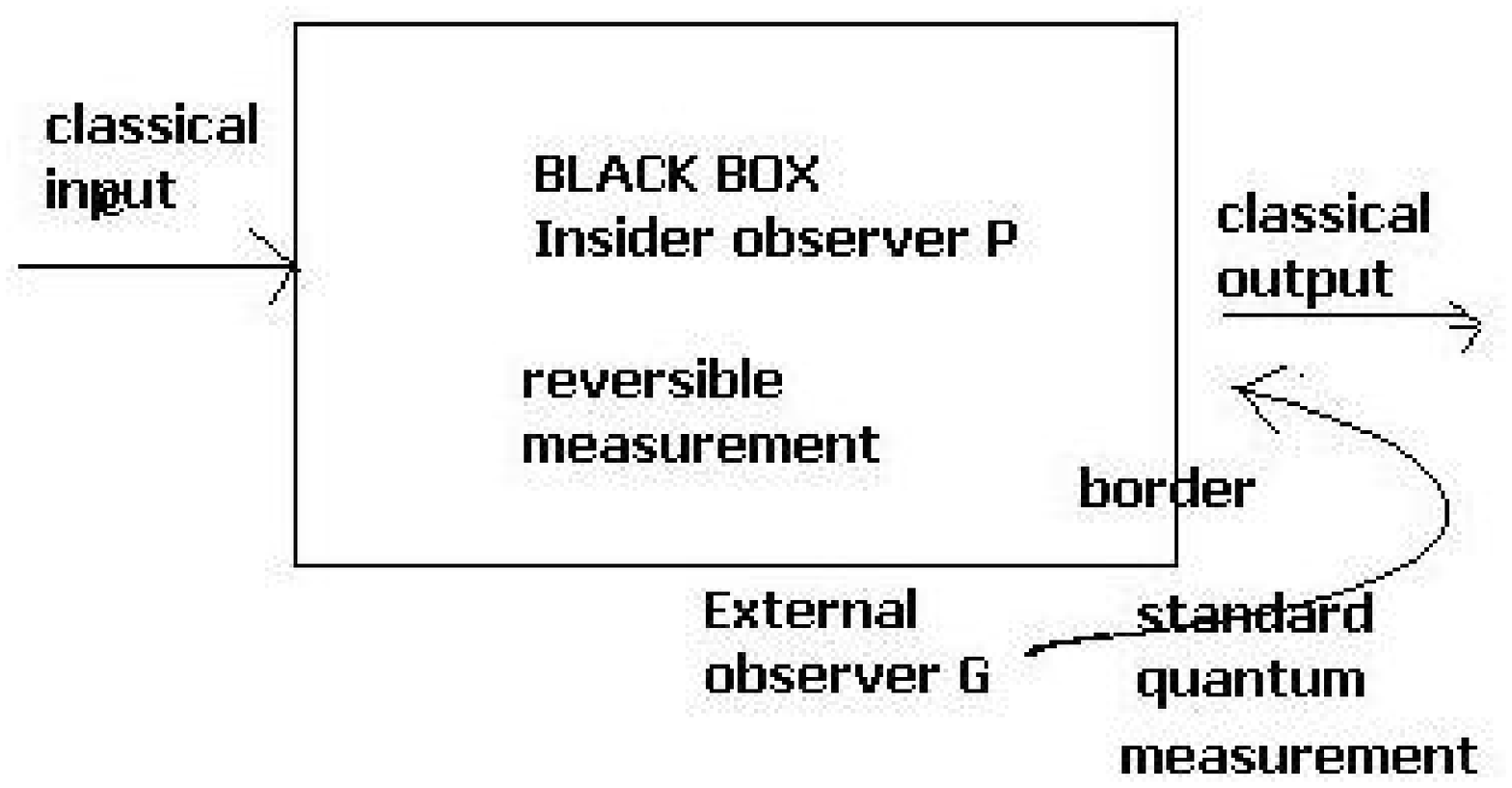}}
\end{figure}
\begin{figure}{{\large
{\bf
Fig. 6}}}
\medbreak
{\large  The Logical Scheme}
\label{figura3}
\centering
\resizebox{12cm}{!}
{
\includegraphics*{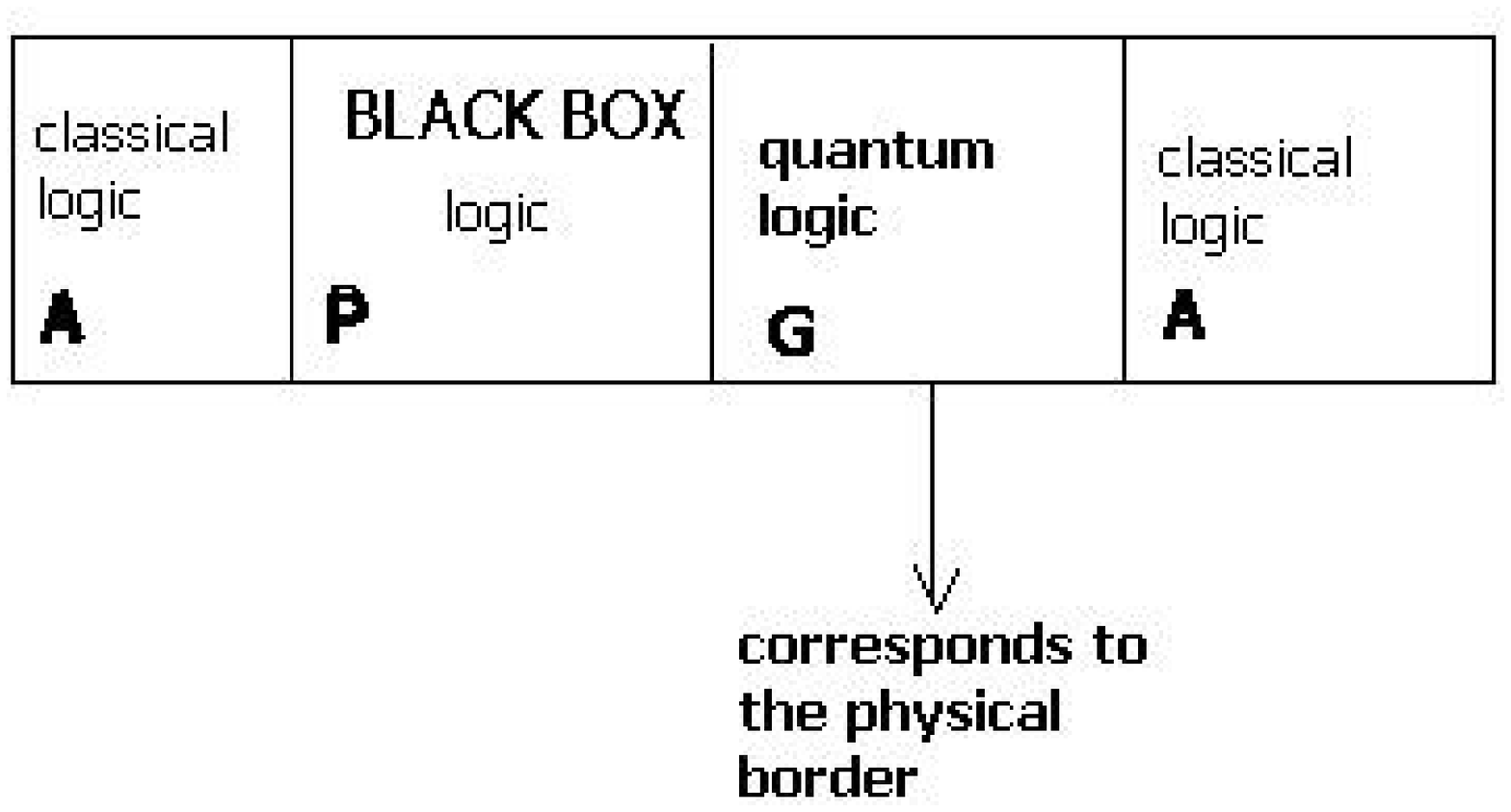}}
\end{figure}
\end{document}